\documentstyle[12pt,psfig,amssymb,amsmath]{article}

\newtheorem{proposition}{Proposition}
\newtheorem{corollary}{Corollary} 
\DeclareMathOperator{\sgn}{\mathrm{sgn}}

\def\1#1{^{(#1)}}
\def\la{\langle}
\def\ra{\rangle}
\begin{document}

\title{On Effective Conductivity on ${\mathbb Z}^d$ Lattice}

\author{Leonid G. Fel${}^{\dag}$ and Konstantin M. Khanin${}^{\ddag}$\\ 
\centerline
{\small {\em
(Dedicated to D. Ruelle and Ya. G. Sinai on occasion of their 65th birthday)}}\\ 
\\${}^{\dag}$
School of Physics and Astronomy 
of Exact Sciences\\Tel Aviv University, Ramat Aviv 69978, Israel\\
and\\
${}^{\ddag}$Isaac Newton Institute for Mathematical Sciences,
\\University of Cambridge,\\
20 Clarkson Road, Cambridge CB3 OEH, UK\\
Heriot-Watt University, Edinburgh\\
Landau Institute, Moscow}

\maketitle
\def\be{\begin{equation}}
\def\ee{\end{equation}}
\def\p{\prime}

\begin{abstract}
We study the effective conductivity $\sigma_e$ for a random wire problem 
on the $d$-dimensional cubic lattice ${\mathbb Z}^d, \, d \geq 2$ in the case
when random conductivities on bonds are independent identically distributed random variables.
We give exact expressions for the expansion of the effective conductivity
in terms of the moments of the disorder parameter up to the 5th order.
In the 2D case using the duality symmetry we also derive the 6th order
expansion. We compare our results with the Bruggeman approximation
and show that in the 2D case it coincides with the exact solution up to the terms
of 4th order but deviates from it for the higher order terms.

\end{abstract}

\vskip 1.cm

\centerline{{\bf Key words:} effective conductivity, Bruggeman's equation}
\newpage

\noindent
\section{Introduction.}
The problem of conductivity of the random composite medium and the
equivalent problem of diffusion in a symmetric (self-adjoint)
random environment has been a subject of intensive study for the
last 25 years. It is virtually impossible to give a full reference list
and we just mention few papers where the mathematical aspects of the
theory were considered for the first time: \cite{kozlov79},\cite{pv81}, \cite{zko81}, 
\cite{khanin82}. In the mathematical literature this problem usually
is quoted as the problem of homogenization for the second order elliptic
differential operators with random coefficients. Roughly speaking
the main result can be formulated in the following way: there exists a non-random effective
conductivity tensor or effective diffusion matrix  such that the
asymptotic properties of the system are the same as for a homogeneous
system governed by the effective parameters. The subject is a very active research area
till now with a vast number of papers publishing every year. However there are very
few results related to the problem of calculation of effective conductivity and
diffusion matrix. In addition to the trivial one-dimensional case such results 
are known only in the self-dual situation in dimension two 
(Keller-Dykhne duality) and in the case of two-component systems where
the analytic continuation method is used to express the effective conductivity
as an analytic function of the ratio of the conductivities of two components
(see \cite{ber}, \cite{mil}, \cite{golpapa}, \cite{bru}, \cite{day}).
In this paper we discuss a very general rigorous method in the lattice case
which was developed in \cite{khanin82}. The method is based on a convergent power series 
expansion for the effective parameters and can be applied for arbitrary probability 
distribution of random conductivities. However, the combinatorics of this expansion is rather 
complicated. That is a reason why it was not used for concrete calculations in the past.

The present paper has two main goals. First of all we demonstrate the constructive
potential of the method in \cite {khanin82} and give exact formulae for the first 5 orders
of the expansion for the effective conductivity in arbitrary dimension.
In the 2D case we also calculate the 6th order terms. We then use our exact results
to study the quality of the classical Bruggeman approximation. We show
that in the 2D case the Bruggeman approximation is extremely accurate
and coincides with the exact answer up to the terms of the 4th order.
We assume everywhere that the random conductivities (jump rates) are independent
identically distributed random variables. Although we consider only the case of 
${\mathbb Z}^d$ lattice we strongly believe that the method can be generalized for other 
types of lattices and even for the continuous situation. 

Yakov Sinai was a teacher of one of us and it is our pleasure to dedicate this 
paper to his 65th birthday. In fact one of the motivations for this paper was to 
illuminate the method developed together with Yakov Grigorevich and to demonstrate 
its effective power.  

\section{Effective conductivity on ${\mathbb Z}^d$ Lattice.}
\subsection{Exact expansion for effective conductivity.}
We consider effective conductivity for a random wire problem on
the $d$-dimensional cubic lattice ${\mathbb Z}^d, \, d \geq 2$. 
Throughout the paper we assume that bond conductivities
$\sigma$ are independent identically distributed positive random variables. 
We are not making any assumptions on a probability distribution of $\sigma$
which can be either discrete or continuous. As we have mentioned above the calculation
of the effective conductivity is equivalent to the calculation of the effective
diffusion matrix for the continuous time random walk in random environment.
In this case random conductivities should be understood as
jump rates through the corresponding bond.
We shall use the formula for the effective diffusion matrix $M_e$ which was
obtained in \cite {khanin82}. This formula is
given by a convergent series where the role of small parameter is played
by a deviation of a random variable $\sigma$ from its average value $\la\sigma \ra.$
Since we consider transitions only along the bonds
of ${\mathbb Z}^d$ lattice with i.i.d. transition rates $\sigma$, the 
effective diffusion matrix is a scalar matrix: 
$M_e = 2\sigma_e I$, where effective diffusion coefficient (or effective
conductivity) $\sigma_e$ can be expressed in terms of a convergent power series.
We first introduce the necessary notations. 

A path 
$\gamma=\{(z_1,\alpha_1), (z_2,\alpha_2), \dots , (z_k,\alpha_k)\}$
is a finite sequence of pairs $(z,\alpha)$ where $z$ is a point of lattice ${\mathbb Z}^d$
and $\alpha = 1,2, \dots , d$ corresponds to one of the $d$ possible directions.
Notice that $z_i,z_{i+1}$ are not necessarily neighbours on the lattice. The sum of
two paths $\gamma = \gamma_1 + \gamma_2$ is simply the ordered union of two sequences
where the pairs of the second path follow the pairs of the first one.
With each pair $(z,\alpha)$ we associate a random variable $\sigma_\alpha(z) =
\sigma (z,z+e_\alpha)$, where $e_\alpha$ is a unit vector in the direction $\alpha$
and $\sigma (z,z+e_\alpha)$ is the random transition rate (conductivity) along
the bond $(z,z+e_\alpha)$. Denote by $u_\alpha(z) = \frac{\sigma_\alpha(z) - \la\sigma\ra}{\la\sigma\ra}$
and define for each path $\gamma=\{(z_1,\alpha_1), (z_2,\alpha_2), \dots , (z_k,\alpha_k)\}$ 
the moment
\begin{equation}
\la\gamma \ra = \la\prod_{i=1}^k u_{\alpha_i}(z_i)\ra \, .
\end{equation}
A convergent expansion below for the effective conductivity is expressed 
through the
moments of a random variable $u$.
We shall also need the following cumulant of a path $\gamma$:
\begin{equation}
E(\gamma)=\sum_{m=1}^k (-1)^{m-1}\sum_{\gamma_1+\dots + \gamma_m=
\gamma}\prod_{j=1}^m\la\gamma_j\ra,
\label{cumulant}
\end{equation}
where summation in (\ref {cumulant}) is taken over all possible partitions of the path
$\gamma$ into a sum of paths $\gamma_j$.  Finally we define a kernel 
$\Gamma_{\alpha \beta}(z)$:
\begin{equation}
\Gamma_{\alpha \beta}(z) = - \int_0^1\dots\int_0^1 
\frac{\sin \pi \lambda_\alpha \sin \pi \lambda_\beta 
\cos 2\pi ((\lambda , z) - \frac{1}{2}\lambda_\alpha + \frac{1}{2}\lambda_\beta)}
{\sum_{\gamma=1}^d \sin^2\pi \lambda_\gamma} \prod_{\gamma=1}^d d\lambda_\gamma \, ,
\end{equation}
where $\lambda = (\lambda_1, \dots , \lambda_d)$. Notice that $\Gamma_{\alpha \alpha}(0) = 
-\frac{1}{d}$ and $\Gamma_{\alpha \beta}(z) = 
\Gamma_{\beta \alpha}(-z)$. We can now write the following
exact formula for $\sigma_e$:
\begin{equation}
\sigma_e = \la\sigma\ra\left(1 + \sum_{k=2}^{\infty}A^{(k)}\right),
\label{exact1}
\end{equation}
where 
\begin{equation}
A^{(k)} = \sum_{\gamma=\{(z_1,\alpha_1), \dots , (z_k,\alpha_k)\} \in {\cal G}_1^{(k)}}
E(\gamma)\prod_{i=1}^{k-1}\Gamma_{\alpha_i \alpha_{i+1}}(z_{i+1} - z_i).
\label {A}
\end{equation}
Here ${\cal G}_1^{(k)}$ is the set of all possible paths $\gamma=\{(z_1,\alpha_1), \dots , 
(z_k,\alpha_k)\}$ such that $z_1 = 0$ and $\alpha_1 = \alpha_d = 1$. It has been proven 
in \cite {khanin82} that the infinite sum in (\ref {A}) is absolutely
convergent. That is due to the fact that for the paths $\gamma$ 
which might lead to divergence of $A^{(k)}$ one has $E(\gamma)=0$. It was also shown that
the expansion in (\ref {exact1}) is absolutely convergent and gives
an exact value of $\sigma_e$ provided $|u| \leq u_0 < 1/2$. The last condition is
technical and probably can be improved. In the following proposition we rewrite 
(\ref {exact1}), (\ref {A}) in a slightly different way.
\begin{proposition} (\cite {khanin82}).

\noindent
Assume that there exists a constant $u_0 < \frac{1}{2}$ such that $|u|\leq u_0$ with
probability 1. Then for any dimension $d$
\begin{equation}
\sigma_e = \la\sigma\ra\left(1 + \sum_{k=2}^{\infty} \ \sum_{m=1}^{\left [\frac{k}{2}\right ]} \ \sum_{s_1, \dots , s_m \geq 2 \atop s_1 + \dots +s_m = k} a_{s_1,\dots ,s_m}\1d \la u^{s_1}\ra \dots  \la u^{s_m}\ra\right) \, ,
\label {exact*}
\end{equation}
where the constants $a_{s_1,\dots ,s_m}\1d$ depend only on dimension $d$ and $[\cdot]$
denotes the integer part. Moreover, for any $n\geq 1$ the following estimate
holds
\begin{equation}
\left |\sigma_e - \la\sigma\ra\left(1 + \sum_{k=2}^{n} \ \sum_{m=1}^{\left [\frac{k}{2}\right ]} \ \sum_{s_1, \dots , s_m \geq 2 \atop s_1 + \dots +s_m = k} a_{s_1,\dots ,s_m}\1d \la u^{s_1}\ra \dots  \la u^{s_m}\ra\right)\right| \leq 
\frac{(2u_0)^{n+1}}{1-2u_0} \, .
\label {exact**}
\end{equation}

\end{proposition}
Note that the series in (\ref {exact*}) is absolutely convergent.

\subsection{The 4th order expansion.}
It is easy to see that only those paths for which each pair $(z,\alpha)$ is present at
least twice give nonzero contribution to (\ref {A}). This immediately
implies that 
\begin{equation}
A^{(2)} = \la  u^2\ra\Gamma_{11}(0)\ = - \frac{\la  u^2\ra}{d}\, , \, A^{(3)} = 
\la  u^3\ra\Gamma_{11}^2(0)= \frac{\la  u^3\ra}{d^2}.
\label {sigma**}
\end{equation}
Hence the 3-rd order approximation to $\sigma_e$ is given by
\begin{equation}
\sigma_e^{(3)} = \la\sigma\ra \left(1 - \frac{\la  u^2\ra}{d} + \frac{\la  u^3\ra}{d^2}\right).
\end{equation}
In the 4-th order the combinatorics is slightly more complicated.
Indeed, nonzero contributions correspond to the paths
\begin{eqnarray}
&&\gamma(4)=\{(0,1),(0,1),(0,1),(0,1)\}\;,\;\;\;\;\;\;\;\;\;\;\;\;\;\;\;\;\;
\gamma^1_{1,z}(4)=\{(0,1),(z,1),(z,1),(0,1)\}, z \neq 0\;,\nonumber \\
&&\gamma_{\alpha,z}(4)=\{(0,1),(z,\alpha),(z,\alpha),(0,1)\}, \alpha \neq 1
\;,\;\;\gamma^2_{1,z}(4)=\{(0,1),(z,1),(0,1),(z,1)\}, z \neq 0\;.\nonumber
\end{eqnarray}
Another possible type of paths $\gamma^3_{1,z}(4)=\{(0,1),(0,1),(z,1),(z,1)\}, z \neq 0$ 
gives zero contribution since $E(\gamma^3_{1,z}(4))=0$. Easy calculation gives
\begin{eqnarray}
A^{(4)} &=& \bigg[(\la  u^4\ra-\la  u^2\ra^2)\Gamma_{11}^3(0) \bigg] + 
\left [\la  u^2\ra^2 \Gamma_{11}(0)\left (\sum_{z \in {\mathbb Z}^d}\Gamma_{11}^2(z) - 
\Gamma_{11}^2(0)\right )\right ]\nonumber \\ 
&+&\left [\la  u^2\ra^2\left (\sum_{z \in {\mathbb Z}^d}\Gamma_{11}^3(z) - 
\Gamma_{11}^3(0)\right )\right ]
+ \sum_{\alpha=2}^d\left [\la  u^2\ra^2\Gamma_{\alpha \alpha}(0)
\sum_{z \in {\mathbb Z}^d}\Gamma_{1\alpha}^2(z)\right ].
\label {sigma}
\end{eqnarray}
Notice that
\begin{equation}
\sum_{z \in {\mathbb Z}^d}\Gamma_{\beta \alpha}^2(z) = 
\int_0^1\dots\int_0^1 \frac{\sin ^2 \pi \lambda_\beta \sin^2 \pi \lambda_\alpha}
{(\sum_{\gamma=1}^d \sin^2\pi \lambda_\gamma)^2} \prod_{\gamma=1}^d d\lambda_\gamma  \, .
\end{equation}
Hence
\begin{equation}
\sum_{\beta , \alpha = 1}^d\sum_{z \in {\mathbb Z}^d}\Gamma_{\beta \alpha}^2(z) = 1 \, .
\end{equation}
Since
\begin{equation}
\sum_{\alpha = 1}^d\sum_{z \in {\mathbb Z}^d}\Gamma_{\beta \alpha}^2(z)
\end{equation}
does not depend on $\beta$ we get
\begin{equation}
\sum_{\alpha = 1}^d\sum_{z \in {\mathbb Z}^d}\Gamma_{1 \alpha}^2(z) = \frac {1}{d} \, .
\label{identity}
\end{equation}
Using (\ref {sigma}, \ref {identity}) we obtain
\begin{equation}
A^{(4)} =
-\frac{1}{d^3}\la  u^4\ra-\frac{d-2}{d^3}\la  u^2\ra^2
+\la  u^2\ra^2\sum_{z \neq 0}\Gamma_{11}^3(z) \, . 
\label{sigma_d}
\end{equation}
The third term in (\ref {sigma_d}) vanishes in the 2D case. 
Indeed, if $z=(x,y)$ we have $\Gamma_{1 1}(x,y) = \Gamma_{22}(y,x)$. Obviously
$\Gamma_{11}(y,x) + \Gamma_{22}(y,x) = 0$ if $(y,x) \neq (0,0)$. Hence, for nonzero 
$(x,y)$ we have $\Gamma_{11}(y,x) = - \Gamma_{1 1}(x,y)$ which immediately implies $\sum_{z \neq 
0}\Gamma_{11}^3(z) = 0$.
As a result we obtain the 4-th order approximation for $d=2$:
\begin{equation}
\sigma_e^{(4)} = \la\sigma\ra\left(1 - \frac{1}{2}\la  u^2\ra + \frac{1}{4}\la  u^3\ra - 
\frac{1}{8}\la  u^4\ra\right) \, .
\end{equation}

\noindent
We next demonstrate that for $d\geq 3$
\begin{equation}
\label{d> 3}
\sum_{z \in {\mathbb Z}^d}\Gamma_{11}^3(z) \neq -\frac{1}{d^3}
\end{equation}
which implies
\begin{equation}
\label{d> 3,1}
\sum_{z \neq 0}\Gamma_{11}^3(z) \neq 0 \, .
\end{equation}
Denote $H(d) = -d^3\sum_{z \in {\mathbb Z}^d}\Gamma_{11}^3(z)$. Using simple Fourier analysis we have
\begin{equation}
\label{d> 3,2}
H(d)= \int_0^1\dots\int_0^1 H(\lambda,\mu)
\prod_{\gamma=1}^d d\lambda_\gamma \prod_{\gamma=1}^d d\mu_\gamma \, , 
\end{equation}
where
\begin{equation}
\label{d> 3,3}
H(\lambda,\mu) = \frac{\sin^2 (\pi(\lambda_1 + \mu_1))}{\frac{1}{d}\sum_{\gamma=1}^d 
\sin^2(\pi(\lambda_\gamma + \mu_\gamma))}\, 
\frac{\sin^2 (\pi \lambda_1)}{\frac{1}{d}\sum_{\gamma=1}^d \sin^2 (\pi \lambda_\gamma)}
\, \frac{\sin^2 (\pi \mu_1)}{\frac{1}{d}\sum_{\gamma=1}^d \sin^2 (\pi \mu_\gamma)} \, .
\end{equation}
As we have explained above the symmetry in the 2D case gives
$\sum_{z \neq 0}\Gamma_{11}^3(z) = 0$ which is equivalent
to $H(2)=1$. We conjecture
that $H(d)$ is a strictly decreasing function of $d$. The conjecture 
implies that  $\sum_{z \neq 0}\Gamma_{11}^3(z) > 0$ for all $d\geq 3$.
Although the conjecture above was not proven rigorously we have checked it numerically
for $3\leq d\leq 5$:
\begin{equation}
H(3)= 0.923 \, , \, H(4) = 0.874 \, , \, H(5) = 0.846 \, .
\end{equation}
Finally, we get the following 4-th order approximation in an arbitrary dimension:
\begin{equation}
\label{sigma_e4d}
\sigma_e^{(4)} = \la\sigma\ra\left (1 - \frac{1}{d}\la  u^2\ra + \frac{1}{d^2}\la  u^3\ra - 
\frac{1}{d^3}\la  u^4\ra - \frac{d+H(d)-3}{d^3}\la  u^2\ra^2\right ) \, .
\end{equation}
\subsection{The 5th order expansion.}
We proceed with the 5-th order calculations. The following paths give nonzero 
contributions:
\begin{eqnarray*}
&&\gamma(5)=\{(0,1),(0,1),(0,1),(0,1),(0,1)\} \, , \, 
\gamma^1_{\alpha,z}(5)=\{(0,1),(z,\alpha),(z,\alpha),(z,\alpha),(0,1)\} \\
&&\gamma^2_{\alpha,z}(5)=\{(0,1),(0,1),(z,\alpha),(z,\alpha),(0,1)\} \, , \, 
\gamma^3_{\alpha,z}(5)=\{(0,1),(z,\alpha),(0,1),(z,\alpha),(0,1)\} \\
&&\gamma^4_{\alpha,z}(5)=\{(0,1),(z,\alpha),(z,\alpha),(0,1),(0,1)\} \, , \, 
{\tilde \gamma}^1_{1,z}(5)=\{(0,1),(z,1),(0,1),(z,1),(z,1)\} \\
&&{\tilde \gamma}^2_{1,z}(5)=\{(0,1),(z,1),(z,1),(0,1),(z,1)\} \, , \, 
{\tilde \gamma}^3_{1,z}(5)=\{(0,1),(z,1),(0,1),(0,1),(z,1)\} \\
&&{\tilde \gamma}^4_{1,z}(5)=\{(0,1),(0,1),(z,1),(0,1),(z,1)\} \, .
\end{eqnarray*}
Notice that in the case $\alpha =1$ the summation in the paths
$\gamma^s_{1,z}(5), {\tilde \gamma}^s_{1,z}(5), 1\leq s \leq 4$
is performed over all $z \neq 0$. Using (\ref {A}) we get 
\begin{equation}
A^{(5)}=\frac{1}{d^4}\la  u^5\ra + K_5(d)\la  u^2\ra\la  u^3\ra,
\end{equation}
where
\begin{equation}
K_5(d) = \sum_{\alpha=1}^d\left( \frac{3}{d^2}
\sum_{z \in {\mathbb Z}^d}\Gamma_{1\alpha}^2(z) 
+\sum_{z \in {\mathbb Z}^d}\Gamma_{1\alpha}^4(z)\right) - \frac{6}{d^4} -
\frac{4}{d}\sum_{z \neq 0}\Gamma_{11}^3(z) \, . 
\end{equation}
This together with (\ref {identity}) gives
\begin{equation}
K_5(d) = \frac{3(d-2)}{d^4}+\sum_{\alpha=1}^d \sum_{z \in {\mathbb Z}^d}
\Gamma_{1\alpha}^4(z) -\frac{4}{d}\sum_{z \neq 0}\Gamma_{11}^3(z) \, . 
\label{K5}
\end{equation}
In the 2D case both the first and the last term in (\ref {K5}) vanish and
\begin{equation}
K_5(2) = \sum_{z \in {\mathbb Z}^2}\Gamma_{11}^4(z) + 
\sum_{z \in {\mathbb Z}^2}\Gamma_{12}^4(z) = I_1 + I_2,
\label{K_5n}
\end{equation}
where
\begin{eqnarray}
\label{I1}
I_1 &=& \int_0^1\int_0^1h_1^2(\lambda_1,\lambda_2)d\lambda_1d\lambda_2 \, , \nonumber \\
h_1(\lambda_1,\lambda_2)&=&\int_0^1\int_0^1\frac{\sin^2 \pi(\lambda_1-\mu_1)
\sin^2 \pi\mu_1 \,d\mu_1d\mu_2}{\bigg(\sin^2 \pi(\lambda_1-\mu_1)+
\sin^2 \pi(\lambda_2-\mu_2)\bigg)\bigg(\sin^2 \pi\mu_1+\sin^2 \pi\mu_2\bigg)}
\end{eqnarray}
and 
\begin{eqnarray}
\label{I2}
I_2 &=& \int_0^1\int_0^1h_2^2(\lambda_1,\lambda_2)d\lambda_1d\lambda_2 \, , \nonumber \\
h_2(\lambda_1,\lambda_2)&=&
\int_0^1\int_0^1\frac{\sin \pi(\lambda_1-\mu_1)\sin \pi(\lambda_2-\mu_2)
\sin \pi \mu_1\sin \pi \mu_2 \, d\mu_1d\mu_2}{\bigg(\sin^2 \pi(\lambda_1-\mu_1)+
\sin^2 \pi(\lambda_2-\mu_2)\bigg)\bigg(\sin^2 \pi\mu_1+\sin^2 \pi\mu_2\bigg)} \, .
\end{eqnarray}
The values of $I_1,I_2$ were found numerically: $I_1 = 0.06391, I_2 = 0.00439$. As a 
result we get in the 2D case the following 5-th order expansion: 
\begin{equation}
\label{sigma_e5}
\sigma_e^{(5)} = \la\sigma\ra\left(1 - \frac{1}{2}\la  u^2\ra + \frac{1}{4}\la  u^3\ra - 
\frac{1}{8}\la  u^4\ra + \frac{1}{16}\la  u^5\ra + 
I\la  u^2\ra\la  u^3\ra\right) \, ,
\end{equation}
where $I = I_1 +I_2 = 0.0683$.

\noindent
In the general case $d\geq3$ we have
\begin{equation}
\sum_{\alpha=1}^d \sum_{z \in {\mathbb Z}^d}\Gamma_{1\alpha}^4(z)=
\sum_{z \in {\mathbb Z}^d}\Gamma_{11}^4(z) + 
\sum_{\alpha=2}^d\sum_{z \in {\mathbb Z}^d}\Gamma_{1\alpha}^4(z)= I_1(d) + (d-1)I_2(d), 
\end{equation}
where
\begin{eqnarray}
\label{I1d}
I_1(d) &=& \int_0^1\dots \int_0^1h_1^2(\lambda )\prod_{\gamma=1}^d d\lambda_\gamma \, , \nonumber \\
h_1(\lambda )&=&\int_0^1\dots \int_0^1\frac{\sin^2 \pi(\lambda_1-\mu_1)
\sin^2 \pi\mu_1 \,\prod_{\gamma=1}^d d\mu_\gamma}{\bigg(\sum_{\gamma=1}^d 
\sin^2(\pi(\lambda_\gamma - \mu_\gamma))\bigg)\bigg(\sum_{\gamma=1}^d \sin^2 (\pi \mu_\gamma)\bigg)}
\end{eqnarray}
and 
\begin{eqnarray}
\label{I2d}
I_2(d) &=& \int_0^1\dots \int_0^1h_2^2(\lambda )\prod_{\gamma=1}^d d\lambda_\gamma \, , \nonumber \\
h_2(\lambda )&=&
\int_0^1\dots \int_0^1\frac{\sin \pi(\lambda_1-\mu_1)\sin \pi(\lambda_2-\mu_2)
\sin \pi \mu_1\sin \pi \mu_2 \, \prod_{\gamma=1}^d d\mu_\gamma}{\bigg(\sum_{\gamma=1}^d 
\sin^2(\pi(\lambda_\gamma - \mu_\gamma))\bigg)\bigg(\sum_{\gamma=1}^d \sin^2 (\pi \mu_\gamma)\bigg)} \, .
\end{eqnarray}
Collecting all the terms we get
\begin{eqnarray}
\label{sigma_e5d}
\sigma_e^{(5)} = \la\sigma\ra\Biggl(1 - \frac{1}{d}\la  u^2\ra + \frac{1}{d^2}\la  u^3\ra - 
\frac{1}{d^3}\la  u^4\ra - \frac{d+H(d)-3}{d^3}\la  u^2\ra^2 \\ \nonumber
+ \frac{1}{d^4}\la  u^5\ra
+ \frac{3d +d^4I(d)+4H(d)-10}{d^4}\la  u^2\ra\la  u^3\ra \Biggr)\, ,
\end{eqnarray}
where $I(d) = I_1(d) + (d-1)I_2(d)$ and $H(d)$ is given by (\ref {d> 3,2}), (\ref {d> 3,3}).

\subsection{Keller-Dykhne duality and the 6th order expansion in the 2D case.}
Although it is possible in principle to calculate an expansion of an
arbitrary order the problem becomes more and more cumbersome for higher
order terms. However in the 2D case one can significantly simplify calculations
using the duality symmetry which was discovered by Keller (\cite {keller64}) and Dykhne
(\cite {dykhne70}). Consider duality transformation
\begin{equation}
\label{dualtransf}
\sigma \rightarrow \frac{1}{\sigma} \ .
\end{equation}
Denote by $\{\sigma \}, \{\sigma^{-1} \}$ the probability
distributions for positive random variables $\sigma$ and $\sigma^{-1}$ respectively.
Then duality symmetry which holds only in the 2D case implies that 
\begin{equation}
\label{sym}
\sigma_e (\{{\sigma^{-1}} \}) = \sigma_e^{-1} (\{\sigma \}) \, .
\end{equation}
Although both Keller and Dykhne considered only the continuous systems the symmetry
(\ref {sym}) can be extended to the case of discrete lattice systems which we study in this paper (see \cite {kogut79}).
The duality symmetry immediately implies that in the self-dual case, i.e. when
the probability distributions  $\{\sigma \}$ and $\{\sigma^{-1} \}$ coincide,
the effective conductivity $\sigma_e = 1$. It also gives an exact answer
in the case which we call almost self-dual. We say that the probability
distribution for a random variable $\sigma$ is almost self-dual
with respect to the duality transformation (\ref {dualtransf}) if
there exists a positive constant $\sigma_0$ such that the probability
distribution for $\sigma_0\sigma$ is exactly self-dual, i.e. 
\begin{equation}
\label{almost}
\{ \sigma_0\sigma\ \}= \{(\sigma_0\sigma )^{-1}\}.
\end{equation}
Since $\sigma_e$ is a homogeneous
function of the first order and $\sigma_e (\{\sigma_0\sigma \})=1$, it follows that in the
almost self-dual situation $\sigma_e (\{\sigma \})=\sigma_0^{-1}$.
Notice that in the two-component case with equipartition, i.e. when $\sigma$ takes values $\sigma_1$ and $\sigma_2$
with probabilities $\frac{1}{2}$ the probability
distribution for $\sigma$ is almost  self-dual with $\sigma_0 = (\sqrt{\sigma_1\sigma_2})^{-1}$.
Hence,
\begin{equation}
\label{dyk}
\sigma_e = \sigma_0^{-1} = \sqrt{\sigma_1\sigma_2} \ .
\end{equation}
This well-known result 
by Keller and Dykhne provides one of the very few exact solutions for the effective
conductivity.

We next show that the duality symmetry alone gives a lot of relations on the
coefficients of the expansion (\ref {exact*}). In fact we shall be able to
recover the 6th order expansion using only the 5th order and the symmetry.
Consider the case when $\sigma$ takes three values: $1-\epsilon$ with probability 
$p$, $1- \alpha \epsilon$ with probability $p$ and $1$ with probability $1-2p$.
Correspondingly a random variable $\sigma^{-1}$ takes values
$\frac{1}{1-\epsilon}$ and $\frac{1}{1- \alpha \epsilon}$ with probabilities $p$
and $1$ with probability $1-2p$. We shall use the formula (\ref {exact*})
in order to calculate
$\sigma_e (\{\sigma \}) \sigma_e (\{\sigma^{-1}\})$ and check the duality identity (\ref {sym})
subsequently in the 2nd, 4th, 6th and 8th orders of the power series expansion in $\epsilon$.
This inductive procedure allows
to find all the relations on the coefficients $a_{s_1,\dots,s_m}\12$.
We performed calculations using the Maple symbolic package. In the 2nd order one
immediately gets $a_2\12=-\frac{1}{2}$. The 4th order calculations give two
relations:
\begin{equation}
\label{rel4}
a_{2,2}\12=\frac{3}{2}a_3\12-\frac{3}{8} \, , \, a_4\12=\frac{1}{4}-\frac{3}{2}a_3\12 \, .
\end{equation}
The 6th order expansion provides four more relations:
\begin{eqnarray}
\label{rel6}
&&a_{2, 2, 2}\12=\frac{7}{2}a_3\12+\frac{3}{2}a_{2,3}\12 - \frac{15}{16} \ , \ 
a_{3,3}\12=\frac{1}{2}+\frac{1}{2}(a_3\12)^2-2a_3\12-a_{2,3}\12 \, , \, \nonumber \\
&&a_{2,4}\12=\frac{11}{8}-6a_3\12 - \frac{3}{2}a_{2,3}\12+ \frac{5}{2}a_5\12 \ , \
a_6\12=\frac{5}{2}a_3\12-\frac{5}{2}a_5\12-\frac{1}{2} \, .
\end{eqnarray}
Using (\ref {sigma_e5}) we have
\begin{equation}
\label{relold}
a_{3}\12=\frac{1}{4} \ , \ a_5\12=\frac{1}{16} \ , \ a_{2,3}\12= I = 0.0683
\end{equation}                           
which immediately gives $a_{2,2}\12=0 \, , \, a_4\12= - \frac{1}{8}$ and
\begin{equation}
\label{rel7}
a_{2, 2, 2}\12= \frac{3}{2}I - \frac{1}{16}
 \ , \  a_{3,3}\12 = \frac{1}{32} - I \ , \  a_{2,4}\12=\frac{1}{32} - \frac{3}{2}I 
  \ , \ a_6\12= - \frac{1}{32} \ .   
\end{equation}
As a result we obtain the 6th order expansion in the 2D case:
\begin{eqnarray}
\label{sigma_e6}
\sigma_e^{(6)}  &=& \la\sigma\ra\Biggr(1 - \frac{1}{2}\la  u^2\ra + \frac{1}{4}\la  u^3\ra - 
\frac{1}{8}\la  u^4\ra + \frac{1}{16}\la  u^5\ra + 
I\la  u^2\ra\la  u^3\ra \nonumber \\
&&- \frac{1}{32}\la  u^6\ra  
- \left(\frac{3}{2}I-\frac{1}{32}\right)\la  u^2\ra\la  u^4\ra
-\left(I-\frac{1}{32}\right)\la  u^3\ra^2 \\
&&+ \left(\frac{3}{2}I-\frac{1}{16}\right)\la  u^2\ra^3 \Biggl) \, . \nonumber
\end{eqnarray}

\section{The Bruggeman Approximation.}
\subsection{Bruggeman's equation.}
The Effective Medium Approximation (EMA) was invented by Bruggeman 
\cite{brug35}, and has remained one of the most popular 
approximations used for calculations of the linear bulk effective electrical
conductivity $\sigma_e$ of a many-component composite medium. This is mainly
due to the simplicity of EMA and to the fact that it gives
accurate results for a wide range of parameters. It also has a
non-trivial percolation threshold which most other simple approximations do
not possess. Another advantage of Bruggeman's approximation is connected
with the fact that none of the complicated details of the microstructure are
used in its construction. EMA is only based on the assumptions that the
composite is macroscopically homogeneous and isotropic and that individual
grains are spherical. It is also important to mention that EMA applies without
any changes to the calculation of dielectric susceptibility, magnetic
permeability, thermal conductivity and chemical diffusion coefficients,
since in all those cases the mathematical structure of the equations  is the
same as for electrical conduction.

Suppose that
the values of the component conductivities $\sigma_i$ and the component volume
fractions $p_i$ are given. Then Bruggeman's equation in the $d-$dimensional case has the following form:
\begin{equation} 
\sum_{i=1}^n p_i\;\frac{\sigma_i-\sigma_B}{\sigma_i+(d-1) \sigma_B}=0\;. 
\label{newt0}
\end{equation} 

>From the mathematical standpoint it has many beautiful properties which are of
high importance for the theory of random composites. 
Equation (\ref{newt0}) has a unique positive root $\sigma_B(\sigma_i)$ which
is homogeneous of the 1-st order, monotone and reducible with respect to 
the equating of some constituents. It is also $S_n$-permutation invariant
in the case when all $p_i$ are equal and compatible with a
trivial solution $\sigma_B =\bar \sigma$ when all $\sigma_i = \bar \sigma \, .$ Finally, in the case $d=2$
the Bruggeman's solution is self-dual with respect to the duality transformation (\ref {dualtransf}).
Namely, if $\sigma_i \rightarrow \sigma_i^{-1}$ and $p_i$ are unchanged then 
\begin{equation}
\sigma_B(\sigma^{-1}_1,\sigma^{-1}_2,...,\sigma^{-1}_n)=\sigma_B^{-1}(\sigma_1,\sigma_2,...,\sigma_n)\;.
\label{keller0}
\end{equation}
It follows that $\sigma_B$ coincides with Keller-Dykhne solutions in the 
self-dual and almost self-dual situations. In particular, $\sigma_B= \sqrt{\sigma_1\sigma_2}$ for the two-component
system with equipartition and conductivities taken values $\sigma_1, \sigma_2$. Notice 
that the Bruggeman approximation is also exact in the $1D$ case.

\subsection{Solution of Bruggeman's Equation.}

Let $\sigma$ be a random variable corresponding to random conductivity.
Then Bruggeman's equation (\ref{newt0}) can be written in terms of averages in the following form
\begin{equation}
\left\la\frac{\sigma-\sigma_B}{\sigma+(d-1)\;\sigma_B}\right\ra=0 \, .
\label{newt1}
\end{equation}
Notice that (\ref {newt1}) is the most general form of Bruggeman's equation. 
We first show that Bruggeman's equation (\ref {newt1}) has a unique positive solution
$\sigma_B$. Indeed, function
\begin{equation}
F(x)=\left\la\frac{\sigma-x}{\sigma+(d-1)\;x}\right\ra
\end{equation} 
is obviously decreasing. Also $F(0)=1$ and $F(x) \to -\frac{1}{d-1}$ as $x \to 
\infty$ which implies the existence and
the uniqueness of the solution. We next find the expansion of $\sigma_B$ in terms of the moments
of the disorder parameter $u=\frac{\sigma-\la\sigma\ra}{\la\sigma\ra}$. It is convenient to introduce
new dimensionless variables 
\begin{equation}
\eta=\frac{\sigma }{\la\sigma\ra}\;,\;\; \xi=\frac{\sigma_B}{\la\sigma\ra}\;\;.
\label{var}
\end{equation}
Obviously $u=\eta -1$. In the new variables Bruggeman's equation (\ref {newt1}) takes the form
\begin{equation}
\left\la\frac{\eta-\xi}{\eta+\delta\;\xi}\right\ra=0\;,
\label{newt1'}
\end{equation}
where $\delta=d-1$. Notice that
\begin{eqnarray}
\frac{\eta-\xi}{\eta+\delta \;\xi}=\frac{1-\xi}{1+\delta\; \xi} +
\frac{(\delta+1)\;\xi\;(\eta -1)}{(1+\delta\; \xi)(\eta + \delta\; \xi)}=
\frac{1-\xi}{1+\delta\;\xi} +\frac{d\;\xi\;u}{(1+\delta\;\xi)^2}\cdot
\sum_{n=0}^{\infty}(-1)^n\left(\frac{u}{1+\delta\;\xi}\right)^n \;.
\label{newt3}
\end{eqnarray}
After the averaging of the both sides in (\ref{newt3}) we get
\begin{equation}
\left\la\frac{\eta-\xi}{\eta+\delta\;\xi}\right\ra=\frac{1-\xi}{1+\delta\;\xi} +\frac{d\;
\xi}{1+\delta\;\xi}\cdot \sum_{n=0}^{\infty}(-1)^n\frac{\la  u^{n+1}\ra}
{(1+\delta\;\xi)^{n+1}} = 0\;
\label{newt4}
\end{equation}
which together with $\la  u\ra=0$ immediately implies
\begin{equation}
\frac{1}{\xi}=1+d\;\sum_{n=2}^{\infty}(-1)^n\frac{\la  u^n\ra}{(1+\delta\;\xi)^n}\;.
\label{newt4a}
\end{equation}
If the random variable $u$ is small enough the solution of equation (\ref {newt4a}) can be written as a convergent
expansion in terms of the moments of $u$:
\begin{equation}
\xi = 1 + \sum_{k=2}^{\infty} \ \sum_{m=1}^{\left [\frac{k}{2}\right ]} \ \sum_{s_1, \dots , s_m \geq 2 \atop s_1 + \dots +s_m = k} b_{s_1,\dots ,s_m}\1d\la  u^{s_1}\ra \dots \la  u^{s_m}\ra \, .
\label {Bexact*}
\end{equation}
Notice that this expansion has similar structure to the expansion (\ref {exact*}). Easy calculation 
leads to the following expansion up to the terms of 6th order:
\begin{eqnarray}
\label{newt6a}
\xi^{(6)} &=& 1 - \frac{1}{d}\la  u^2\ra + \frac{1}{d^2}\la  u^3\ra - 
\frac{1}{d^3}\la  u^4\ra - \frac{d-2}{d^3}\la  u^2\ra^2 + \frac{1}{d^4}\la  u^5\ra \nonumber\\
&&+ \frac{3d - 5}{d^4}\la  u^2\ra\la  u^3\ra -\frac{1}{d^5}\la  u^6\ra-\frac{4d-6}{d^5}\la  u^2\ra\la  u^4\ra \\
&&-\frac{2d-3}{d^5}\la  u^3\ra^2-\frac{2d^2-8d+7}{d^5}\la  u^2\ra^3 \, , \nonumber
\end{eqnarray}
which gives the 6th order approximation for the Bruggeman approximation 
\begin{equation}
\label{brug6}
\sigma_{B}^{(6)}=\la\sigma\ra\xi^{(6)}
\end{equation}
and its 2D version
\begin{eqnarray}
\label{brug6a}
\sigma_{B}^{(6)}  = \la\sigma\ra\Biggr(1 - \frac{1}{2}\la  u^2\ra + \frac{1}{4}\la  u^3\ra - 
\frac{1}{8}\la  u^4\ra + \frac{1}{16}\la  u^5\ra + \frac{1}{16}\la  u^2\ra\la  u^3\ra \nonumber \\
- \frac{1}{32}\la  u^6\ra  -\frac{1}{16}\la  u^2\ra\la  u^4\ra-\frac{1}{32}\la  u^3\ra^2 
+ \frac{1}{32}\la  u^2\ra^3 \Biggl) \, . 
\end{eqnarray}
\subsection{Effective conductivity and the Bruggeman approximation.}

It follows from (\ref {sigma_e5d}), (\ref {newt6a}), (\ref {brug6}) that the Bruggeman approximation $\sigma_B$
coincides with the effective conductivity $\sigma_e$ up to the terms of 3rd order. However if
$d \geq 3$ the 4th order terms are different. Let us assume that
\begin{equation}
\label{assumpeps}
|u|\leq \epsilon \ , \ \la  u^2\ra \ \geq c\epsilon^2 \ .
\end{equation}
Then,
\begin{equation}
\label{diff4d}
\sigma_e - \sigma_B = \la\sigma\ra\left(\frac{1-H(d)}{d^3}\la  u^2\ra^2 + O(\epsilon^5)\right) \geq \
\la\sigma\ra\left(\frac{1-H(d)}{d^3}c^2\epsilon^4 + O(\epsilon^5)\right)\ .
\end{equation} 
This implies that for $\epsilon$ small enough $\sigma_e > \sigma_B$. In the 2D
case the Bruggeman approximation is even more accurate. It coincides with
$\sigma_e$ up to the 4th order terms. Nevertheless, if $\la\sigma^3\ra$ does not vanish
then $\sigma_e$ differs from $\sigma_B$ in the 5th order.  Assume that (\ref {assumpeps}) holds and in
addition $|\la  u^3\ra| \geq c\epsilon^3$. Then,
\begin{equation}
\label{diff52}
\sigma_e - \sigma_B = \la\sigma\ra\left(\left(I -\frac{1}{16}\right)\la  u^2\ra\la  u^3\ra + O(\epsilon^6)\right) \ .
\end{equation}
Since $I > \frac{1}{16}$ we have $\sigma_e \neq \sigma_B$ for $\epsilon$ small enough.
Notice that $\sigma_B$ is bigger than $\sigma_e$ if $\la  u^3\ra$ is negative.
Finally we consider the symmetric 2D case. We shall assume that $u$
satisfies (\ref {assumpeps}) and $\la  u^3\ra=0$. Then,
\begin{equation}
\label{diff62}
\sigma_e - \sigma_B = \la\sigma\ra\left(\frac{3}{2}\left(\frac{1}{16}-I\right)\la  u^2\ra\la(u^2-\la  u^2\ra)^2\ra + O(\epsilon^7)\right) \ .
\end{equation} 
It follows from (\ref {diff62}) that $\sigma_e < \sigma_B$ if $\la(u^2-\la  u^2\ra)^2\ra$ 
is of the order of $\epsilon^4$ and $\epsilon$ is small enough. We summarise all
three cases in the following simple proposition.
\begin{proposition} \

1. Consider the case $d \geq 3$. If $u$ satisfies (\ref {assumpeps}) then 
there exists $\epsilon(d,c) > 0$ such that $\sigma_e > \sigma_B$ for all $\epsilon \leq \epsilon(d,c)$.

2. Let $d=2$, $u$ satisfies (\ref {assumpeps}) and $|\la  u^3\ra| \geq c\epsilon^3$.
Then there exists $\epsilon(c) > 0$ such that $\sigma_e \neq \sigma_B$ for all $\epsilon \leq \epsilon(d,c)$
and $\sgn{(\sigma_e - \sigma_B)} = \sgn{(\la  u^3\ra)}$.

3. Let $d=2$ and $\la  u^3\ra=0$. If $u$ satisfies (\ref {assumpeps}) and $\la(u^2-\la  u^2\ra)^2\ra \ \geq c\epsilon^4$
then there exists ${\bar {\epsilon}}(c) > 0$ such that $\sigma_e < \sigma_B$ for all $\epsilon \leq {\bar {\epsilon}}(c)$.
\end{proposition}

\noindent
The following corollary follows easily from Proposition 2. Consider the
$n-$component system where $\sigma$ takes the values $\sigma_1, \sigma_2, \dots,
\sigma_n$ with probabilities $p_1, p_2, \dots, p_n$, \, 
$p_i> 0$, $p_1+p_2+\dots +p_n=1$. We shall also assume that the system is irreducible,
i.e. $\sigma_i \neq \sigma_j, \, 1\leq i,j \leq n$. Denote $p_{\rm min}=\min(p_1, p_2, \dots, p_n)$.
\begin{corollary} \

1. Let $d\geq 3$. If $|u| \leq \epsilon(d,p_{\rm min})$ then $\sigma_e > \sigma_B$. 

2. Let $d=2$. Assume that $n=2$ and $p_1 > p_2$. Then $\sigma_e \neq \sigma_B$
provided  $|u| \leq \epsilon(c)$, where $c=p_2\left(1-(\frac{p_2}{p_1}\right)^2)$.
Moreover, if $\sigma_2 > \sigma_1$ then $\sigma_e > \sigma_B$. In the opposite case,
i.e. when $\sigma_1 > \sigma_2$ one has $\sigma_B > \sigma_e$.

3. Let $d=2$. Assume that $n=3$ and $\la  u^3\ra \, = 0$. Then there exists
$c_1(p_1,p_2,p_3)> 0$ such that $\sigma_e < \sigma_B$
if $|u| \leq {\bar {\epsilon}}(c_1)$.
In particular, if $\sigma_1 = 1+ \epsilon, \sigma_2 = 1, \sigma_3 = 1- \epsilon$
and $p_1=p_3=p, \,  p_2=1-2p, \, 0 < p < \frac{1}{2}$ then $c_1(p_1,p_2,p_3)=2p(1-2p)$
and $\sigma_e < \sigma_B$ under condition
$|u| \leq {\bar {\epsilon}}(2p(1-2p))$.
\end{corollary}

Finally, we conjecture that
for $n-$component systems the effective conductivity coincides with the Bruggeman
approximation only if the probability distribution $\{\sigma \}$ is almost self-dual,
see (\ref {almost}).

\section{Concluding remarks.}

\noindent
1. We have derived the exact formulae for the first 5 orders of the
expansion of the effective conductivity in terms of the moments of the disorder
parameter $u$ in arbitrary dimension. In the 2D case we have also found the 6th
order terms. It is quite interesting to extend these results to other types of 
2D lattices and to the continuous plaquetes systems. Notice that our duality
analysis holds for the general 2D case. Hence, if the expansion (\ref {exact*})
is valid, it is enough to find $a_{3}\12, a_{5}\12, a_{2,3}\12$ in order to
determine all other terms up to the 6th order.

\noindent
2. We have shown that Bruggeman's solution (\ref {brug6a}) gives
a remarkably accurate approximation for the effective conductivity
of the 2D random many-component lattice wire system. It turns out
that in the case of square lattice the first four orders of the
expansion of Bruggeman's solution in terms of the moments of the disorder 
parameter coincide with the corresponding expansion of the exact solution. 
However, in the 5th order the Bruggeman approximation deviates from the exact 
one. An interesting  and natural problem is to verify whether such behaviour is characteristic for 
the square lattice or it also holds for other 2D lattices.
It is also interesting to analyse the relation between Bruggeman's solution 
and effective conductivity for the continuous 2D random composites. 
Recently four isotropic 
three-component $S_3$-permutation invariant regular structures with three-fold 
rotation lattice symmetries in the 2D case were treated numerically \cite{felberg20}. 
A simple cubic equation with one free parameter $A\geq 0$
$$
\sigma_e^3+A\; J_1 \sigma_e^2-A\; J_2 \sigma_e-J_3=0\;,\;\;\;\;
J_1=\sum_{i=1}^3\sigma_i\;,\;J_2=\sum_{i\neq j}\sigma_i \sigma_j\;,\;
J_3=\sigma_1 \sigma_2 \sigma_3\;
$$
was proposed as an algebraic equation of minimal order. Its solution share many 
properties with $\sigma_e$ and corresponds to Bruggeman's solution 
when $A=\frac{1}{3}$. The numerically estimated values of $A$ corresponding to different cases 
were calculated with a very high precision. It appears that they are 
distinct and lie rather far from $\frac{1}{3}$ for some of the structures.
This indicates a strong dependence of 
$\sigma_e$ on plane symmetries in contrast with the two-component case.

\noindent
3. Recently
in the paper by A. Kamenshchik and I. Khalatnikov (\cite{Khalat20}) the perturbation theory was developed
for the periodic three-component plaquetes lattice systems with two-fold rotation lattice symmetry. We hope
that their technique combined with our approach will lead to the exact expansion
for the effective conductivity in the random plaquetes situation.

\noindent
4. After the paper was submitted we were informed about the paper by
Jean-Marc Luck (\cite{luck}) where very similar results were obtained
using different method for calculating an expansion for the effective 
conductivity. In our opinion the approach we use has certain
advantages. First of all, it is rigorous and, hence, more suitable
for mathematical audience. Secondly, it gives arbitrary good rigorous bounds
for the effective conductivity (see (\ref {exact**})).
\section{Acknowledgment.}

The authors would like to thank A. Kamenshchik and I. Khalatnikov for useful 
discussions. We are also grateful to Jean-Marc Luck for bringing
his paper (\cite {luck}) to our attention and to D. Khmelev for his help
with numerical calculations.

\noindent
The main part of the paper has been written during the stay of one of the authors (LGF) at the Isaac 
Newton Institute for Mathematical Sciences and its hospitality is highly 
appreciated.

\noindent
The work was supported in part by grants from the Tel Aviv University
Research Authority and the Gileadi Fellowship program.


\end{document}